\documentclass[
    ,final            
  ]
  {aipproc}

\layoutstyle{6x9}

\newcommand{\NP}[1]{{ Nucl.\ Phys.\ } {\bf  #1}}
\newcommand{\ZP}[1]{{ Z.\ Phys.\ } {\bf  #1}}

\newcommand{\PL}[1]{{ Phys.\ Lett.\ } {\bf  #1}}

\newcommand{\PR}[1]{{ Phys.\ Rev.\ } {\bf  #1}}
\newcommand{\PRL}[1]{{ Phys.\ Rev.\ Lett.\ } {\bf  #1}}

\newcommand{\lsim}{\raise.3ex\hbox{$<$\kern-.75em\lower1ex\hbox{$\sim$}}}
\newcommand{\ima}{{\mbox{Im}\,}}
\newcommand{\rea}{{\mbox{Re}\,}}
\newcommand{\be}{\begin{equation}}
\newcommand{\ee}{\end{equation}}

\begin{document}

\title{Poles of light resonances from
unitarized Chiral perturbation Theory
and their large $N_c$ behavior
\footnote{Invited talk to The 25$^{th}$ Annual Montreal-Rochester-Syracuse-Toronto Conference on High Energy Physics (``J. Schechter Fest''). 
May 13 - 15,  2003, Syracuse University,
Syracuse, New York  }}

\author{J.R.Pel\'aez}{address={Dipartimento di Fisica. 
Universita' degli Studi and INFN,
 Firenze,
  Italy},
  altaddress={Departamento de F\'{\i}sica Te\'orica II, 
  Universidad Complutense, 28040 Madrid, Spain}
}

\begin{abstract}
We have recently completed the one loop
calculation of meson-meson scattering within Chiral Perturbation Theory.
Once unitarized, these amplitudes provide simultaneously 
a remarkable description of the resonance region up to 1.2 GeV 
as well as the low energy region,
respecting the chiral symmetry expansion.
The position of the poles in these amplitudes
is related to the mass and width of the associated resonances
that are generated without being explicitly included in the Lagrangian.
The spectroscopic nature of these states 
can then be deduced by studying the
behavior of these poles, through the $N_c$ dependence
of the Chiral Perturbation Theory parameters, which 
can be obtained from QCD in the large $N_c$ limit.
\end{abstract}

\maketitle

I take advantage of the occasion to celebrate 
the long and fruitful physics carrier of J. Schechter,
to whom this workshop is dedicated in his 65th anniversary,
to review my work, either recently published or still in progress,
which is closely related 
to some of Joe's relevant contributions.

For many years now, we take QCD as the fundamental theory
of strong interactions. Its predictions have been 
thoroughly tested to 
great accuracy
in the perturbative regime (above 1-2 GeV), where a description
in terms of quarks and gluons is possible. 
However, QCD becomes non-perturbative 
at low energies and the usual perturbative expansion has 
to be abandoned
in favor of somewhat less systematic approaches in terms of 
mesons. An exception
is the formalism of Chiral Perturbation Theory (ChPT) \cite{chpt1,chpt2}, 
built as the most general derivative expansion of a Lagrangian
containing pions, kaons and the eta. These particles 
are identified as the Goldstone bosons associated to the spontaneous
chiral symmetry breaking of the massless QCD theory.
In terms of observables ChPT becomes an expansion in
powers of energy or momenta over a scale $4\pi f_0\simeq 1.2\,$GeV.
For the processes we are interested in, and due to Lorentz invariance,
only even powers of energy and momenta occur in the expansion,
which are generically denoted as $O(p^2), O(p^4)$, etc...
The quark masses, small compared with typical hadronic scales,
are introduced in the same power counting, and give rise to the 
masses of the $\pi, K$ and $\eta$ mesons, counted as $M=O(p^2)$. 
The ChPT Lagrangian
thus allows for true  Quantum Field Theory calculations, including
meson loops, whose divergences are renormalized in a set of
chiral parameters
at each order in the expansion.
In particular, the theory is finite at each order
and predictive in the sense that once the set of parameters
up to a given order is determined from some experiments, these 
very same parameters can, {\it and should} describe, to that order,
 any other 
physical process involving mesons. As usual in a renormalization
program, the ``bare'' parameters appearing in the Lagrangian
may depend on an arbitrary 
 regularization scale $\mu$;  however, the physical observables
are scale independent, since the $\mu$ dependence is canceled
through the regularization of the loop integrals. In other words,
all the relevant physical scales are those given by parameters
in the Lagrangian. Another salient feature of ChPT
is its model independence and the fact that it has been
possible to relate its parameters to those of QCD, 
at least in the limit of a large number of colors $N_c$
\cite{chpt2,chptlargen}.

Unfortunately, the applicability of ChPT
is limited to low energies or momenta. 
First, because
the number of independent terms allowed by symmetry
  increases dramatically  at each order.
Second, because resonances appear rather soon
in meson physics and these states are
associated to poles in the second Riemann sheet of the amplitudes.
Such a non-perturbative behavior cannot be accommodated by truncated series
as those of ChPT (as a matter of fact there are also logarithmic terms,
but for this discussion, only the powers count). A third  way of seeing
this problem is the violation of the unitarity constraint,
which becomes more and more severe as the energy increases.

During the last few years, unitarization methods have emerged
as a powerful tool to extend ChPT
to the resonant region 
\cite{Dobado:1996ps,Oller:1997ng,Guerrero:1998ei,GomezNicola:2001as}. The basic point is to realize that 
the unitarity of 
meson-meson partial waves, $t$,  determines completely the 
imaginary part of the inverse of the amplitude $\ima 1/t$. Actually, the 
dynamical information comes through $\rea 1/t$, which has to be calculated
from theory. The use of ChPT  to
calculate $\rea 1/t$ has yielded remarkable results. 
In particular, we have recently shown \cite{GomezNicola:2001as} that,
by unitarizing the one-loop ChPT amplitudes,
it is possible to generate the
resonant behavior of light meson resonances,
while respecting at the same time the low energy chiral expansion
(in particular the values of the chiral parameters already determined
from other processes).

In what follows we will review all those results
and present our recent determination \cite{Coimbra} of the
pole positions of the generated resonances, which are related to
their masses and widths. We will also present
an study of the behavior of these poles in the large $N_c$ limit.
This whole approach is of special interest
 for the meson spectroscopy community, since these resonances are
generated from the most general Lagrangian consistent with QCD,
and therefore without any bias toward its existence.
Let us recall that in particular the existence of light scalar
mesons is still under a strong debate, and the appearance
of these states in our work could help shedding some light in this issue.
The fact that nine of these scalar poles seem to appear together,
suggests that they could form a multiplet.
Finally, if the existence of these states is rather controversial,
even more so is
their nature or composition in terms of quarks and gluons.
However, quarks and gluon states are properly defined only in a 
microscopic  
framework, and our approach allows for a clear link with QCD
in the large $N_c$ limit, where these states have a consistent
definition.

\section{Meson meson scattering within Unitarized Chiral Perturbation Theory}
Customarily partial waves $t_{IJ}$ of
definite isospin $I$ and angular momentum $J$
are used  to compare with experiment. Omitting for simplicity 
the $I,J$ subindices, the chiral expansion becomes
$t\simeq t_2+t_4+...$, with $t_2$ and $t_4$ of ${\cal
O}(p^2)$ and ${\cal O}(p^4)$, respectively. 
The unitarity relation  is
rather simple for the partial waves $t_{ij}$,
where $i,j$ denote the different available states. For
instance, when two states, say "1" and ``2'', are accessible,
it becomes
\be
\ima T = T \, \Sigma \, T^* \quad \Rightarrow \quad \ima T^{-1}=- \Sigma
\quad  \Rightarrow \quad T=(\rea T^{-1}- i \,\Sigma)^{-1}
\label{unimatrix}
\ee
with \vspace*{-.5cm}
\be
T=\left(
\begin{array}{cc}
t_{11}&t_{12}\\
t_{12}&t_{22} \\
\end{array}
\right)
\quad ,\quad
\Sigma=\left(
\begin{array}{cc}
\sigma_1&0\\
0 & \sigma_2\\
\end{array}
\right)\,,
\ee
where $\sigma_i=2 q_i/\sqrt{s}$ and $q_i$ is the C.M. momentum of
the state $i$. Let us remark again that
{\it we only need to know the real part of the Inverse Amplitude}
since the imaginary part is fixed by unitarity.
Note that the
unitarity relations are non-linear and therefore cannot be satisfied 
exactly with a perturbative expansion like that of ChPT. 
Nevertheless,
unitarity holds perturbatively, i.e,
\vspace*{-.2cm}
\begin{eqnarray}
\ima T_2 = 0, \quad \quad \ima T_4 = T_2 \, \Sigma
\, T_2^*\,+ {\cal O}(p^6) . \label{pertuni}
\end{eqnarray}
Using in eq.(\ref{unimatrix})
the ChPT expansion of 
$\rea T^{-1}\simeq  T_2^{-1}(1-(\rea T_4) T_2^{-1}+...)$ 
ensures that we will be respecting the ChPT low energy expansion
thus and taking into account all the information
included in the chiral Lagrangians.

Let us recall that the leading order Lagrangian is universal
since it only depends on the chiral symmetry breaking scale $f_0$,
which corresponds to the pion decay constant in the chiral limit,
and the leading order meson masses $M^0_\pi, M^0_K$ and $M^0_\eta$.
The dependence on the QCD dynamics only comes through the one loop
chiral parameters, which for meson-meson scattering in SU(3) are 
called $L_i$, with $i=1...8$ \cite{chpt2}.

As a matter of fact there is an additional simplification that
is allowed in ChPT, 
since the $O(p^6)$ and higher orders 
in Eq.({\ref{pertuni}) can be made to vanish
if  the substitution of $f_0$ 
in terms of the observables $f_\pi$ or $f_K$ or $f_\eta$
is made to match their corresponding powers on both
sides of the above equations. In such case:
\vspace*{-.2cm}
\begin{eqnarray}
\ima T_2 = 0, \quad \quad \ima T_4 = T_2 \, \Sigma
\, T_2^*. \label{exactpertuni}
\end{eqnarray}
We will call these conditions ``exact perturbative unitarity''.
From  eq.(\ref{exactpertuni}),
we find
\begin{equation}
 T\simeq T_2 (T_2-T_4)^{-1} T_2,
\label{IAM}
\end{equation}
which is the coupled channel IAM, 
which we have used 
to unitarize simultaneously all the one-loop ChPT meson-meson
scattering amplitudes \cite{GomezNicola:2001as}.

Part of this unitarization program had been carried out first for 
partial waves in the elastic region, where a single channel
is enough to describe the data. In such a way were found 
the $\rho$ and $\sigma$ poles
in $\pi\pi$ scattering and that of  $K^*$ in $\pi K\rightarrow\pi K$
\cite{Dobado:1996ps}. 
The $\kappa$ pole can also be
obtained in the elastic single channel formalism but was only noticed
later in this context.
A simplified coupled channel calculation, considering 
only the leading order and the dominant s-channel loops, but 
neglecting crossed and tadpole loop
diagrams, was used in \cite{Oller:1997ng}, since at that time
not all the meson-meson amplitudes were known to one-loop. 
Despite the approximations, it resulted in
a remarkable description of meson-meson
scattering up to 1.2 GeV. The poles associated to the 
$\rho$, $K^*$, $f_0$, $a_0$, $\sigma$ and $\kappa$
resonances were found 
when these partial waves were continued to
the second Riemann sheet of the complex energy plane.
However, the approximations implied that only 
the leading order of the expansion
was recovered at low energies. In addition,
the divergences were regularized with a cutoff, which
could violate chiral symmetry  by introducing an spurious parameter.
Furthermore, due to this cutoff regularization,
it was not possible to compare the $L_i$ parameters,
which are supposed to encode the underlying QCD dynamics, 
with those already present in the literature,
obtained from other low energy processes.
This, together with a 
residual cutoff dependence due to the
absence of the whole set of one loop diagrams makes it
hard to study the large $N_c$ behavior that, as we will see
is inherited by ChPT from QCD

Since not only the nature, but also the existence of 
light scalar resonances is rather controversial,
it is very relevant to check that
these poles and their features
are not just artifacts of the approximations,
estimate the uncertainties in their parameters, and
check their compatibility with other experimental information
regarding ChPT.

That is why the $K\bar{K}\rightarrow K\bar{K}$ one-loop amplitudes
were calculated in \cite{Guerrero:1998ei}: together with previous
workss \cite{Kpi}, the \cite{Guerrero:1998ei} results
allowed for the unitarization of the $\pi\pi$, $K\bar{K}$
coupled system. They found good agreement
of the IAM description with the existing $L_i$, reproducing
again the resonances in that system.
Finally, we have
completed more recently \cite{GomezNicola:2001as}
the one-loop meson-meson scattering calculation.
There are three new amplitudes:
$K\eta\rightarrow K\eta$, $\eta\eta\rightarrow\eta\eta$ and
$K\pi\rightarrow K\eta$, but we have recalculated
the other five amplitudes unifying the  notation, ensuring
`` exact perturbative
unitarity'', Eq.(\ref{exactpertuni}), 
and also correcting some errors in the literature.
Next, we have applied the coupled channel IAM  to these amplitudes.
Our results allow for a direct comparison with the
standard low-energy chiral parameters, which we find in very good
agreement with previous determinations from low-energy data. 
The main differences with \cite{Oller:1997ng}
are: i) we consider the full calculation of all the one-loop
amplitudes in dimensional regularization, ii) we are able to describe
simultaneously  the low energy and the resonance regions, and iii) 
we pay special attention to the estimation of uncertainties.

First of all we checked that the resonant features were still present
with the standard ChPT parameters, that we have given in Table 1.
As already commented, this comparison can only be performed now since
we have all the amplitudes
renormalized in the standard $\overline{MS}-1$ scheme.

In Fig.1 we show the results.  Let us recall that
meson-meson scattering data are very hard to obtain, 
and usually have large systematic uncertainties,
which are the largest contribution to the
error bands of the curves
in the figure as well as in the parameters in Table 1.
As it happened
in \cite{GomezNicola:2001as}, the uncertainty bands
are calculated from a MonteCarlo Gaussian sampling (1000 points)
of the $L_i$ sets within their error bars, assuming they are
uncorrelated.

As a matter of fact in Table 1 we give three sets of
parameters for the IAM, reasonably compatible among them and with standard
ChPT parameters.
The IAM I fit was obtained reexpressing  all the $f_0$
appearing in the amplitudes by their expression
in terms of just $f_\pi$,
which. although simpler,  is somewhat unnatural 
when dealing with kaons or etas.  
The plots and the uncertainties of this fit
were already given in \cite{GomezNicola:2001as} and therefore
the corresponding plots are not shown here.
There, it could be observed that the description
of the $f_0(980)$  region was
somewhat poor, yielding
a too small width for the resonance.

That is the reason why  
in Fig.1 we present the  results \cite{Coimbra} using
amplitudes written in terms of 
$f_K$ and $f_\eta$ when dealing with processes
involving kaons or etas, that we call set IAM II.
As it can be noticed, the data in the $f_0(980)$ region
is well within the uncertainties.
In particular, we have rewritten
our $O(p^2)$ amplitudes changing one factor of $1/f_\pi$ by
$1/f_K$ for each two kaons present
between the initial or final state, or by $1/f_\eta$ 
for each two etas appearing between the initial and final states.
In the special case $K\eta\rightarrow K\pi$ we have changed 
$1/f_\pi^2$ by $1/(f_Kf_\eta)$.
Of course, these changes introduce some corrections at $O(p^4)$
which can be easily obtained using the relations between
the decay constants and $f_0$ provided in \cite{chpt2,GomezNicola:2001as}. 
The $1/f_\pi$ factor in each loop function  at $O(p^4)$
(generically, the $J(s)$
given in the appendix of \cite{GomezNicola:2001as})
have to be changed according to eqs.(\ref{exactpertuni}),
that is, they satisfy the ``exact perturbative unitarity''
condition and are the same as 
the previous ones, up to $O(p^6)$ differences.
However, at high energies there can be some
 numerical differences.
Note that the only parameters that
suffer a sizable change are those related to the definition of
decay constants: $L_4$ and $L_5$. 
We also give in Table 1 a third fit, IAM III, obtained as IAM II
but fixing $L_4$ to zero as in 
the most recent $K_{l4}$  $O(p^4)$ 
determinations. This is the value 
estimated from the large $N_c$ limit,
and since our fits are not very sensitive to the 
variations in $L_4$ in this
way we avoid that it could get an unnatural value 
that correspond just to a very tiny improvement
of the $\chi^2$.
\begin{table}[hbpt]
\begin{tabular}{|c||c|c||c|c|c|}
\hline
  \tablehead{1}{r}{b}{Parameter} &
  \tablehead{1}{c}{b}{
$K_{l4}$ decays} &
  \tablehead{1}{c}{b}{\hspace*{0.6cm}ChPT\hspace*{0.6cm} } &
  \tablehead{1}{c}{b}{\hspace*{0.6cm}IAM I\hspace*{0.6cm}}&  
  \tablehead{1}{c}{b}{\hspace*{0.6cm}IAM II\hspace*{0.6cm}} &
  \tablehead{1}{c}{b}{\hspace*{0.6cm}IAM III\hspace*{0.6cm}} 
\\
\hline
$L_1^r(M_\rho)$
& $0.46$
& $0.4\pm0.3$
& $0.56\pm0.10$ 
& $0.59\pm0.08$
& $0.60\pm0.09$
\\
$L_2^r(M_\rho)$
& $1.49$
& $1.35\pm0.3$ 
& $1.21\pm0.10$ 
& $1.18\pm0.10$
& $1.22\pm0.08$\\
$L_3 $  &
 $-3.18$ &
 $-3.5\pm1.1$&
$-2.79\pm0.14$ 
&$-2.93\pm0.10$
& $-3.02\pm0.06$
\\
$L_4^r(M_\rho)$
& 0 (fixed)
& $-0.3\pm0.5$& $-0.36\pm0.17$ 
& $0.2\pm0.004$
& 0 (fixed)\\
$L_5^r(M_\rho)$
& $1.46$
& $1.4\pm0.5$& $1.4\pm0.5$ 
& $1.8\pm0.08$
& $1.9\pm0.03$
\\
$L_6^r(M_\rho)$
& 0 (fixed)
& $-0.2\pm0.3$& $0.07\pm0.08$ 
&$0\pm0.5$
&$-0.07\pm0.20$\\
$L_7 $  & $-0.49$ & 
$-0.4\pm0.2$&
$-0.44\pm0.15$ &
$-0.12\pm0.16$&
$-0.25\pm0.18$
\\
$L_8^r(M_\rho)$
& $1.00$
& $0.9\pm0.3$& $0.78\pm0.18$ 
&$0.78\pm0.7$
&$0.84\pm0.23$\\
\hline
\end{tabular}
\caption{Different sets of chiral parameters ($\times10^{3}$).
The first column comes from recent analysis of $K_{l4}$ decays
\cite{BijnensKl4} ($L_4$ and $L_6$ are set to
zero). In the ChPT column $L_1,L_2,L_3$ come from
\cite{BijnensGasser} and  the rest from  \cite{chpt2}. The 
three last 
ones correspond to the values from the IAM including the
uncertainty due to different systematic error used on different
fits. Sets II and III are obtained using amplitudes 
expressed in terms of $f_\pi$, $f_K$ and $f_\eta$, whereas
the amplitudes in set I are expressed in terms of $f_\pi$ only.} \label{eleschpt}
\end{table}
\begin{table}[h]
\begin{tabular}{|c|c|c|c|c|}
\hline 
Threshold&Experiment
&\hspace{0.5cm}IAM fit I\hspace*{0.5cm}&\hspace{0.5cm}ChPT ${\cal O}(p^4)\hspace{0.5cm}$
&\hspace*{0.5cm}ChPT ${\cal O}(p^6)$\hspace*{0.5cm}\\
parameter&&\cite{GomezNicola:2001as}&\cite{Dobado:1996ps,Kpi}
&\cite{Amoros:2000mc}\\ 
\hline \hline 
$a_{0\,0}$&0.26 $\pm$0.05&0.231$^{+0.003}_{-0.006}$&0.20&0.219$\pm$0.005\\
$b_{0\,0}$&0.25 $\pm$0.03&0.30$\pm$ 0.01&0.26&0.279$\pm$0.011\\
$a_{2\,0}$&-0.028$\pm$0.012&-0.0411$^{+0.0009}_{-0.001}$&-0.042&-0.042$\pm$0.01\\
$b_{2\,0}$&-0.082$\pm$0.008&-0.074$\pm$0.001&-0.070&-0.0756$\pm$0.0021\\
$a_{1\,1}$&0.038$\pm$0.002&0.0377$\pm$0.0007&0.037&0.0378$\pm$0.0021\\ 
$a_{1/2\,0}$&0.13...0.24&0.11$^{+0.06}_{-0.09}$&0.17&\\
$a_{3/2\,0}$&-0.13...-0.05&-0.049$^{+0.002}_{-0.003}$&-0.5&\\
$a_{1/2\,1}$&0.017...0.018&0.016$\pm$0.002&0.014&\\
$a_{1\,0}$&&0.15$^{+0.07}_{-0.11}$&0.0072&\\ 
\hline
\end{tabular}
\caption{ Scattering lengths $a_{I\,J}$ and slope parameters
$b_{I\,J}$ for different meson-meson scattering channels. For
experimental references see \cite{GomezNicola:2001as}. Let us
remark that our one-loop IAM results at threshold are very similar
to those of two-loop ChPT.}\label{elesfit}
\end{table}

\begin{figure}[hbpt]
\includegraphics[height=0.85\textheight]{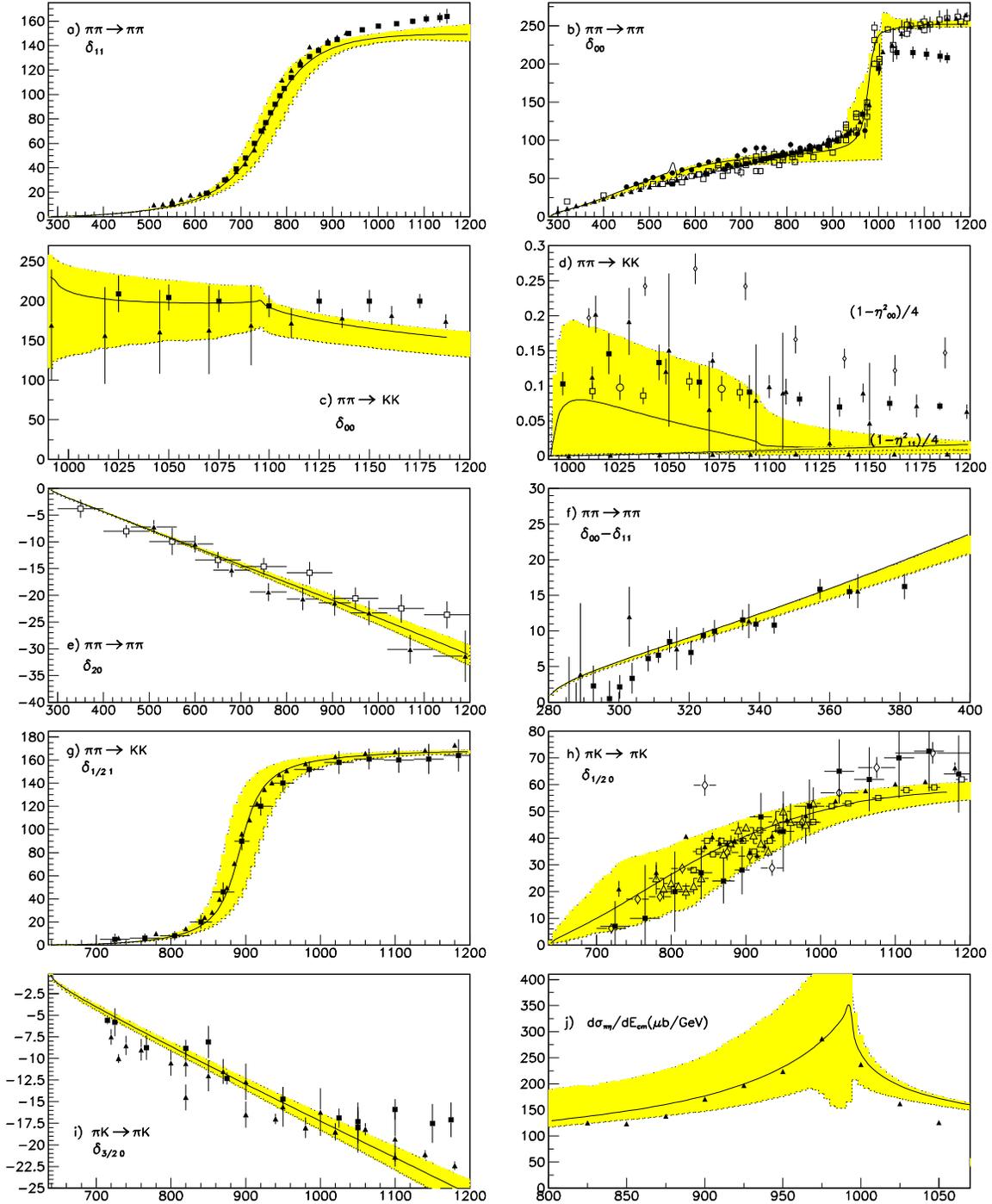}
\caption
{ IAM fit to meson-meson scattering data, set II in Table 1.
The uncertainties cover  the estimated systematic errors.
The statistical errors from the fit would be much smaller.
The data comes from \cite{pipidata}} 
\label{fig:Tps1}
\end{figure}

In conclusion, we have a set of amplitudes obtained from unitarized
ChPT that describe meson-meson scattering 
data up to energies of 1.2 GeV. That is, they describe the resonant
region while respecting simultaneously the low energy constraints,
ensuring a remarkable low energy description, as can be seen from the 
values of the the scattering 
lengths and slopes listed in Table 2. In addition, we have checked
that the chiral parameters needed for this unitarized description
are compatible with those from standard ChPT, and, since the expressions
are fully renormalized, we are not including any dependence on any  
spurious parameter.

\section{T-matrix poles associated to resonances}

Thus, with the amplitudes just commented we will now study whether
we still obtain the same poles as obtained with previous approaches.
The  $\sigma$ and $\kappa$ states are still rather controversial, 
although their poles have been found by several groups (that of Joe Schechter
among others) with different approaches \cite{newsigma}. In addition,
recent experiments seem to require such poles \cite{charm}, 
supporting strongly their existence since 
they have completely different systematics
compared with meson scattering processes.

The poles in unitarized chiral amplitudes
have triggered the interest in the hadron spectroscopy
community since they are not included in the ChPT Lagrangian
and are therefore generated without any theoretical bias toward
their existence, classification in multiplets, or nature.
Since the scalar resonances $\sigma, \kappa, a_0(980)$ and $f_0(980)$
appear together in these chiral symmetric approaches, it seems very
natural to interpret that they belong to the same nonet.
Nevertheless, we should distinguish two types of resonances:
already in \cite{Oller:1997ti} it was noted that the scalars
could be generated with just a cutoff, that is, without the need to include
one-loop chiral parameters, in contrast with the vector mesons,
which needed these parameters, particularly $L_1, L_2$ and $L_3$
\cite{Oller:1997ng}.
The vectors seem to be fairly well established as $q\bar{q}$ states,
and this difference hints that the lightest scalar nonet
may have a different nature. With the amplitudes described in the previous
section we expect to reach a more conclusive statement,
since they respect the chiral expansion, and, being completely renormalized,
have no
spurious dependence on any cutoff or 
dimensional regularization scale.

Thus, we show in Table 3 the poles and their uncertainties,
for the different IAM sets
of parameters given in Table 1, compared with those obtained
in the ``approximated'' IAM \cite{Oller:1997ng}.
For illustration we show in Table 4 the poles listed presently in the
PDG \cite{PDG}. Let us comment briefly these results for the different resonances:
\begin{itemize}
\item The vectors $\rho(770)$ and $K^*(892)$, are very stable
in the chiral unitary approaches. They appear consistently 
in single channel approaches \cite{Dobado:1996ps}, 
approximated coupled channel IAM \cite{Oller:1997ng}
and the complete IAM \cite{GomezNicola:2001as}.

\item The $\sigma$ and $\kappa$ poles are robust
within these approaches. They appear with compatible values
in single channel approaches, approximated coupled channel IAM 
and the complete IAM. Note the small uncertainties in some of their parameters.

\item  The $f_0(980)$  mass is robust,
and the pole appears both in the approximated and complete IAM approaches
(of course it cannot appear in the single channel case since it has a sizable
 decay to two different channels). However, we have already remarked
that the unitarized ChPT amplitudes using just $f_\pi$
as those given in \cite{GomezNicola:2001as}, yield a somewhat poor description
of its width. Nevertheless is well described when using $f_\pi$,
$f_K$ and $f_\eta$.

\item   Although the data in the $a_0(980)$ region are always well
described, the presence of a pole is strongly dependent on how 
the unitarized ChPT amplitudes are parametrized. There 
is no
pole when using only $f_\pi$, which was 
already pointed out in \cite{Uehara:2002nv},  where, using the
``approximated'' coupled channel IAM with just $f_\pi$,
a ``cusp'' interpretation
of the $a_0(980)$ enhancement in $\pi\eta$ production was suggested. However,
 we do find a pole 
when using $f_\pi$,
$f_K$ and $f_\eta$ either with the approximated or complete IAM. 
Thus, this pole is rather unstable as can 
be noticed from its large uncertainties in Table 3. We are 
somewhat more
favorable toward the pole interpretation because the use of $f_\pi$,
$f_K$ and $f_\eta$ is more natural and is also able to 
describe better the $f_0(980)$ width.

\end{itemize}

Let us nevertheless note that the $f_0(980)$ and $a_0(980)$ resonances
are very close to the $K\bar{K}$ threshold, and they couple strongly to this 
state. The proximity of this threshold can produce a considerable distortion
in the resonance shape, whose relation to the pole position
can then be far from the expected one for narrow resonances. In addition
these states have a large mass ad it is likely that their nature 
should be understood from a mixture with heavier states.

\begin{table}[htbp]
\begin{tabular}{ccccccc}
\hline
$\sqrt{s_{pole}}$(MeV)
&$\rho$
&$K^*$
&$\sigma$
&$f_0$
&$a_0$
&$\kappa$
\\ \hline
$^{\hbox{IAM Approx}}_{\;\;\;
\hbox{(no errors)}}$
&759-i\,71
&892-i\,21
&442-i\,227
&994-i\,14
&1055-i\,21
&770-i\,250
\\\hline
IAM I
&760-i\,82
&886-i\,21
&443-i\,217
&988-i\,4
& cusp?
&750-i\,226
\\
(errors)
&$\pm$ 52$\pm$ i\,25
&$\pm$ 50$\pm$ i\,8
&$\pm$ 17$\pm$ i\,12
&$\pm$ 19$\pm$ i\,3
&
&$\pm$18$\pm$i\,11
\\ \hline
IAM II
&754-i\,74
&889-i\,24
&440-i\,212
&973-i\,11
&1117-i\,12
&753-i\,235
\\
(errors)
&$\pm$ 18$\pm$ i\,10
&$\pm$ 13$\pm$ i\,4
&$\pm$ 8$\pm$ i\,15
&$^{+39}_{-127}$ $^{+i\,189}_{-i\,11}$
&$^{+24}_{-320}$ $^{+i\,43}_{-i\,12}$
&$\pm$ 52$\pm$ i\,33\\\hline
IAM III
&748-i68
&889-i23
&440-i216
&972-i8
&1091-i52
&754-i230
\\
(errors)
&$\pm$ 31$\pm$ i\,29
&$\pm$ 22$\pm$ i\,8
&$\pm$ 7$\pm$ i\,18
&$^{+21}_{-56}$$\pm$ i\,7
&$^{+19}_{-45}$ $^{+i\,21}_{-i\,40}$
&$\pm$ 22$\pm$ i\,27\\\hline
\end{tabular}
\caption{ Pole positions (with errors) in meson-meson scattering.
When close to the real axis the mass and width of the 
associated resonance is $\sqrt{s_{pole}}\simeq M-i \Gamma/2$.}
\end{table}

\begin{table}[htbp]
\begin{tabular}{ccccccc}
\hline
PDG2002
&$\rho(770)$
&$K^*(892)^\pm$
&$\sigma$ or $f_0(600)$
&$f_0(980)$
&$a_0(980)$
&$\kappa$
\\ \hline
Mass (MeV)
&$771\pm0.7$
&$891.66\pm0.26$
&(400-1200)-i\,(300-500)
&$980\pm10$
&$980\pm10$
&not\\
Width (MeV)
& $149\pm0.9$
& $50.8\pm0.9$
&(we list the pole)
&40-100
&50-100
&listed\\\hline
\end{tabular}
\caption{ Mass and widths or pole positions 
of the light resonances quoted in the PDG \cite{PDG}.
Recall that for narrow resonances $\sqrt{s_{pole}}\simeq M-i \Gamma/2$}
\end{table}

\section{Chiral Perturbation Theory and the large $N_c$ }

QCD is not perturbative at low energies, say below 1 or 2 GeV.
In this region, however, the limit of a large number of colors $N_c$,
despite $N_c$ being only 3 in nature,
has emerged as a powerful tool to understand
many qualitative features of QCD and
also as a guiding line to organize calculations \cite{largen}.
This is one more topic that Joe has addressed in his long career \cite{largen}.
The advantage of the large $N_c$ limit to study resonances is that 
$q\bar{q}$ states become real bound states as $N_c\rightarrow\infty$. 
In particular,
it has been shown that their mass should be basically constant
$M\simeq O(1)$, whereas
their decay width to two mesons should decrease as $\Gamma\simeq O(1/N_c)$.
A similar behavior should hold for glueballs decaying to two mesons.
In contrast, some multiquark states as $qq\bar{q}\bar{q}$ are expected to 
become simply unbound, that is, the meson-meson continuum \cite{Jaffe}.

ChPT inherits the large $N_c$ features of QCD
through 
the meson masses and decay constants, that behave as 
$f_\pi, f_K, f_\eta=O(\sqrt{N_c})$
and $M_\pi,M_K,M_\eta=O(1)$, and also 
through the scaling of the chiral parameters \cite{chptlargen},
 that we list in Table 5.
There are also estimates of the value for $N_c=3$, also listed 
in Table 5. One must bear in mind that there is an uncertainty
on the renormalization scale where these estimates should be evaluated,
although on general grounds one expects them to be valid for 
$\mu\simeq 0.5-1\,$GeV.
From Table 5 we see that, despite
3 is not
such a large number and the scale uncertainty, 
the agreement with both ChPT and the IAM parameters
is fairly impressive.

\begin{table}[hbpt]
\begin{tabular}{|c||c||c||c||c|}
\hline
  \tablehead{1}{r}{b}{Parameter\\
$\times 10^{-3}$} &
  \tablehead{1}{c}{b}{ChPT\\
\hspace*{.5cm}$\mu=770$ MeV\hspace*{.5cm}} &
  \tablehead{1}{c}{b}{IAM I\\
\hspace*{.5cm}$\mu=770$ MeV\hspace*{.5cm}}&  
  \tablehead{1}{c}{b}{\hspace*{.5cm}Large $N_c$,\hspace*{.5cm}\\
$N_c=3$} &
  \tablehead{1}{c}{b}{Large $N_c$\\ \hspace*{.5cm}behavior\hspace*{.5cm}} 
\\
\hline
$2 L_1- L_2$
& $-0.6\pm0.2$
& $0.56\pm0.10$ 
& $0$
& $O(1)$
\\
$L_2$
& $1.35\pm0.3$ 
& $1.21\pm0.10$ 
& $1.8$
& $O(N_c)$\\
$L_3 $  &
 $-3.5\pm1.1$&
$-2.79\pm0.14$ 
&$-4.3$
& $O(N_c)$
\\
$L_4$
& $-0.3\pm0.5$& $-0.36\pm0.17$ 
& $0$
& $O(1)$\\
$L_5$
& $1.4\pm0.5$& $1.4\pm0.5$ 
& $2.1$
& $O(N_c)$
\\
$L_6$
& $-0.2\pm0.3$& $0.07\pm0.08$ 
&$0$
&$O(1)$\\
$L_7 $  &
$-0.4\pm0.2$&
$-0.44\pm0.15$ &
$-0.3$&
$O(1)$
\\
$L_8$
& $0.9\pm0.3$& $0.78\pm0.18$ 
&$0.8$
&$O(N_c)$\\
\hline
\end{tabular}
\caption{Different sets of chiral parameters ($\times10^{3}$).
For illustration, the 
ChPT and IAM I columns are repeated from Table 1. 
Other IAM sets give similar results. In the third
column we give the leading large $N_c$ estimates for the
chiral parameters, with $N_c$ set to 3. 
The last column shows the leading large $N_c$ behavior,
calculated from QCD. 
} 
\label{elesln}
\end{table}

\section{The large $N_c$ behavior of the lightest resonances}

\begin{figure}[hbpt]
\includegraphics[height=0.45\textheight]{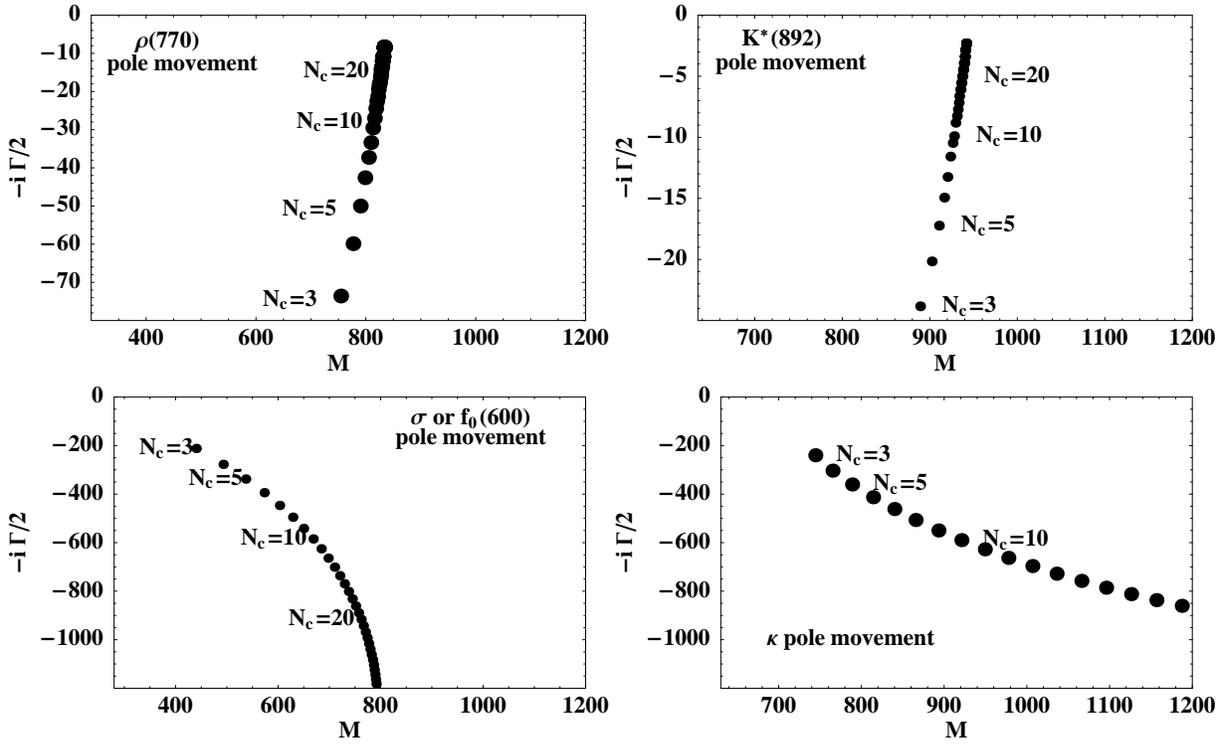}
\caption
{ Large $N_c$ dependence of the pole positions
in the lower half of the second Riemann $\sqrt{s}$ 
sheet of the meson meson scattering amplitude, obtained
from the unitarized one-loop Chiral Perturbation Theory calculation.
For each value of $N_c$ the pole is represented by a dot,
in different meson-meson
scattering channel. 
Note that the $\sigma$ and $\kappa$ behavior is opposite to that
of well know vector states as the $\rho$ and $K^*$.}  
\label{fig:largen}
\end{figure}

The fact that ChPT is the most general Lagrangian compatible
with the QCD chiral symmetry breaking, and in particular
that it inherits the large $N_c$ features of QCD, will
allow us to study the large $N_c$ behavior of these resonances
and get a hint on their nature. We will simply impose
the scaling of the ChPT parameters at $\mu=770\,$MeV  in the IAM amplitudes
fitted to the data, which therefore correspond to $N_c=3$.
For {\it narrow} resonances the pole position
is related to the mass and width of a given resonance by 
$\sqrt{s_{pole}}\simeq M-i\Gamma/2$. By comparing the actual behavior
of the poles with that expected from large $N_c$ 
we could, in principle, elucidate the nature (or at least the dominant
component) of the resonant states.
We will next show the results of our, still preliminary, study.

In Fig.2 we thus show the displacement of the 
poles in several channels of meson meson scattering.
Each dot corresponds to a different value of $N_c$.
First of all, we want to test the method, and for
that we turn to vector resonances, which are well
established $q\bar{q}$ states.
Thus, on the top left, we represent the $I,J=1,1$ channel of $\pi\pi$
scattering and thus the movement
of the $\rho(770)$ pole as $N_c$ increases. On the top right we display
the same but for $\pi K$ for $I,J=1,1/2$, that is, the $K^*(982)$ movement.
Remarkably, in both cases we can notice 
that their mass tends to a constant and that 
the width decreases as $1/N_c$, as expected for $q\bar{q}$ states.

Once we have checked that this approach reproduces
correctly the expected behavior of  $q\bar{q}$ states, we turn
to the controversial scalar resonances.
On the bottom left of Fig.2 we represent the movement of the pole 
commonly associated to the $\sigma$ and on the right, that associated with
the $\kappa$ resonance in $\pi K$ scattering. 
For simplicity we keep the notation
$\sqrt{s_{pole}}\simeq M-i\Gamma/2$, although
 these poles are so far from the real axis 
that the interpretation in terms of mass and width
is no longer straightforward. 
Note that the behavior of both the $\sigma$ and $\kappa$ poles
is completely at variance with that expected for $q\bar{q}$ states.
In particular, both $M$ and $\Gamma$ grow with $N_c$ for both states.
This behavior rules out both the $q\bar{q}$ and glueball interpretations.
Let us remark, however, that the four-quark state interpretation
(also two-meson molecules) \cite{Jaffe}
is able to accomodate the fact that these states become
the meson-meson continuum as $N_c$ grows.

\section{Conclusion and outlook}

We have shown that unitarized meson-meson scattering
ChPT amplitudes provide a simultaneous description
of the low energy and resonant regions below 1.2 GeV.
This description respects the chiral symmetry constraints,
and in particular the low energy chiral expansion up to 
next to leading order. It is also renormalization
scale independent, thus avoiding any spurious dependence from
artificial scales.
We have seen that the unitarized
fit leads to 
parameters compatible with those of standard ChPT
and that it yields scattering lengths compatible with
higher order calculations and the most recent low energy data.

With these amplitudes we have shown that 
all the resonant shapes are reproduced, and that their
description in terms of poles is robust, with the possible
exception of the $a_0$ (that could alternatively be interpreted as a cusp).
We have provided uncertainties for the determinations 
of these poles. Since our amplitudes are built from
chiral symmetry and unitarity in a very general way,
those states which are robust under different unitarization
techniques, seem to be an unavoidable requirement of chiral symmetry
and unitarity.

We have also presented an on-going study of the large $N_c$
behavior of the poles generated with the IAM and ChPT.
The behavior of the vectors follows remarkably well the
expected behavior, given its established $q\bar{q}$ nature.
In contrast, the $\sigma$ and $\kappa$ states, behave
in a completely different way, which disfavors a
$q\bar{q}$ or glueball interpretation. A $qq\bar{q}\bar{q}$
interpretation is qualitatively adequate to explain the observed
behavior.

However, as already remarked, this study is still preliminary.
At present, we are finishing the study of the 
large $N_c$ behavior of the $f_0$ and $a_0$ poles,
for which the situation is more cumbersome due
to the larger uncertainty in the $a_0$ interpretation from
unitarized amplitudes,
the proximity of thresholds and possible mixings with
more massive multiplets 
\footnote{While preparing this work, an
study of the mass matrix of scalar resonances 
that survive in the large $N_c$ has appeared \cite{Cirigliano:2003yq}.
Claiming that the $f_0$ could be such a state and 
also the $a_0$ in one possible scenario}. We are also
estimating the errors due to the uncertainty in the 
renormalization scale where the large $N_c$ scaling is imposed.
Finally, the interpretation of wide
resonances is somewhat fishy in terms of poles
and for that reason we are presently looking directly
at the large $N_c$
behavior of the amplitude (the phase shift) in the physical region.
The results will be presented elswhere soon \cite{prep}, hoping
they could be of use in
clarifying the spectroscopic status of light meson scalars
in the near future.

\begin{theacknowledgments}
First of all, I wish to thank the MRST'03 organizers, 
and specially A. Fariborz, for their  kind invitation 
and for their efforts to offer us such a pleasant and lively workshop
in Syracuse. In addition, 
I wish to thank A. G\'omez Nicola for his comments and his careful reading 
of the manuscript. I also thank A. Andrianov, E. Espriu, F. Kleefeld 
for encouraging me to look for the large $N_c$ behavior of the poles, as well
as R. Jaffe for his comments on the large $N_c$ 
behavior of multiquark hadrons.
Work supported by a Marie Curie fellowship, contract MCFI-2001-01155,
the Spanish CICYT projects, FPA2000-0956,
PB98-0782 and BFM2000-1326, the
CICYT-INFN collaboration grant 003P 640.15, and the 
E.U. EURIDICE network contract no. HPRN-CT-2002-00311.
\end{theacknowledgments}

\bibliographystyle{aipproc}   

\end{document}